\documentclass[a4paper,fleqn,usenatbib]{mnras}

\usepackage[T1]{fontenc}
\usepackage{ae,aecompl}
\usepackage[utf8]{inputenc}
\usepackage{color}

\usepackage{graphicx}   
\usepackage{amsmath}    
\usepackage{amssymb}    
\usepackage{float}

\title[Synthetic indices for multiple populations in GCs]
{Synthetic spectroscopic indices for identifying multiple stellar populations in globular clusters}

\author[E. Bertone, M. Chavez, and J. C. Mendoza]
{ Emanuele Bertone \thanks{E-mail: ebertone@inaoep.mx},
Miguel Chávez \thanks{mchavez@inaoep.mx},
J. César Mendoza \thanks{cesar.mendoza@inaoep.mx}
\\
Instituto Nacional de Astrof{\'\i}sica, \'Optica y Electr\'onica, Luis Enrique Erro 1, CP 72840, Tonantzintla, Puebla, Mexico
}

\date{Accepted XXX. Received YYY; in original form ZZZ}

\pubyear{2020}

\begin{document}

\label{firstpage}
\pagerange{\pageref{firstpage}--\pageref{lastpage}}
\maketitle

\begin{abstract}
We present an investigation of synthetic spectroscopic indices that can plausibly help in identifying the
presence of multiple stellar populations in globular clusters. The study is based on a new grid of stellar model atmospheres and high-resolution ($R$=500,000) synthetic spectra, that consider chemical partitions that have been singled out in Galactic globular clusters. The database is composed of 3472 model atmospheres and theoretical spectra calculated with the collection of Fortran codes DFSYNTHE, ATLAS9 and SYNTHE, developed by Robert L. Kurucz. They cover a range of effective temperature from 4300 to 7000~K, surface gravity from 2.0 to 5.0~dex and four different chemical compositions. A set of 19 spectroscopic indices were calculated from a degraded version ($R$=2500) of the theoretical spectra dataset. The set includes five indices previously used in the context of globular clusters analyses and 14 indices that we have newly defined by maximizing the capability of differentiating the chemical compositions. We explored the effects of atmospheric parameters on the index values and identified the optimal spectral diagnostics that allow to trace the signatures of objects of different 
stellar populations, located in the main sequence, the red giant branch and the horizontal branch. We found a suitable set of indices, that mostly involve molecular bands (in particular NH, but also CH and CN), that are very promising for spectroscopically identifying multiple stellar populations in globular clusters.
\end{abstract}

\begin{keywords}
globular clusters: general -- stars: abundances -- stars: main sequence -- stars: red giant branch -- stars: horizontal branch -- techniques: spectroscopic
\end{keywords}



\section{Introduction}

Until recently, globular clusters (GCs) were considered as stellar systems hosting a single generation of stars which are initially chemically homogeneous. Such a concept has been used as a workhorse in the analysis of more complex stellar populations, as galaxies in the nearby and distant universe \citep{Bressan1994}. Nevertheless, early studies of GCs already demonstrated that some objects showed significant dispersion in the chemical abundance of their members \citep[][and references therein]{Osborn1971,Freeman1975}, as well as quite different morphologies of the color magnitude diagram (CMD) at advanced evolutionary stages such as the horizontal branch (HB) \citep[e.g.][]{Castellani2005, Caloi2008, DAntona2008, Catelan2009, Dalessandro2013, Valcarce2016}. Currently, it is well known that many Galactic GCs are made up of multiple stellar populations, with a more complex star-formation history than that suggested by the simple stellar population concept. Most of Galactic GCs (perhaps all) appear to have two (or more) approximately coeval subpopulations formed through different star-formation phases, which causes the observed peculiar chemical patterns.

Before the multiple stellar population were photometrically discovered, a wealth of works devoted to chemical studies of GCs had been conducted, starting with that of \citet{Osborn1971} who laid the groundwork for future studies. In that investigation, he reported chemical abundance anticorrelations between the strength of cyanide (CN) and methylidyne (CH) bands in M5 and M10. This first evidence of chemical inhomogeneity was detected in studies of the more easily accessible stars, i.e., the stars of the bright red giant branch (RGB). Other subsequent works also focused in RGB stars and similarly reported unusual, for a single generation of stars, chemical patterns that showed a bimodal distribution of the strength of CN \citep[e.g.][]{Norris1981,Norris1981solo,Smith1982}.

More recently, many other GCs have been studied, some of which also incorporated stars on the main sequence (MS) \citep{Kayser2008,Pancino2010} and showed that the anticorrelation between CN and CH bands is also a feature in non-evolved objects. In addition to the CN-CH abundance anticorrelation, similar trends for oxygen and sodium and for magnesium and aluminum have been found. For instance, high resolution studies on stars of 47~Tucanae by \citet{Carretta2004} and of M3 and M13 by \citet{Cohen2005}, followed by the analysis of hundreds of objects of 19 southern GCs in a range of metallicities (-2.4$\leq$[Fe/H]$\leq$-0.4~dex) by \citet{Carretta2009, Carretta2010}, demonstrated that Na-O and Mg-Al anticorrelations are also ubiquitous in GCs and that such trends are more evident in metal-poor and/or massive clusters \citep[see][]{Johnson2005,Bragaglia2015,Villanova2017}.

Since the early investigations, a number of explanations have been brandished to explain the observed abundance anomalies \citep[see, e.g., the reviews of][]{Gratton2004,Gratton2012} and, in view of the evolved nature of the studied stars in most works, the first proposed scenario involved chemical atmospheric enrichment due to mixing during the first dredge-up along the RGB, in which the products of the CNO cycle would be brought to the surface. However, while mixing could partially explain the observed chemistry, the variation of C and N abundances cannot be explained in MS stars, where the CNO cycle does not dominate. It has also been proposed that the chemical anomalies must be already present in the pristine material from which stars were formed and that a polluted second stellar generation (Na-richer and O-poorer) can explain the observed chemical patterns. Still, none of the scenario presented up to now may consistently explain the whole collection of observational data.

The helium content plays an important role in the identification of stellar population in GCs. In fact, it has been suggested that the GCs showing a wide difference in the abundance of He are not an exception but the rule, i.e., it is the most common result of the GC formation process \citep{DAntona2008}. Stars showing chemical anomalies also present He enrichment \citep{Piotto2007, Caloi2008, DAntona2008}. An higher He content has a strong effect on advanced evolutionary stages such as the HB \citep[e.g.][]{Castellani2005, Caloi2008, DAntona2008, Catelan2009, Dalessandro2013, Valcarce2016}. An explanation is that, while a star loses mass, it increases its He abundance and therefore He-enriched objects populate bluer regions of the HB \citep{DAntona2008}. The blue HB stars with He enhancement ought to be brighter than the red HB stars without He enhancement \citep{Catelan2009}.

The first direct testimony of the presence of multiple stellar populations in GCs was provided by the photometric studies of $\omega$ Centauri \citep{Lee1999, Bedin2004, Villanova2007, Bellini2010}, one of the first GCs that showed a significant variation in the chemical composition of its member stars and an anomalously wide RGB \citep{Freeman1975}. Being the most massive GC of the Milky Way ($\sim 3.9\times10^6$ M$_{\odot}$) \citep{Pryor1993}, it was thought to be a unique and peculiar case. Nevertheless, to date, and thanks to the exquisite photometric data collected by 8-10 m class telescopes and the Hubble Space Telescope, it is now evident that most of GCs possess two or more stellar populations, with some striking examples as M2, that shows as many as seven different populations \citep{Milone2015}.

Theoretical stellar atmospheres and spectra have been used in several works to explore the effects of chemical mixtures representative of different stellar populations of GCs on photometric colors. 
For instance, \citet{Marino2008} computed the synthetic Johnson $(U-B)$ color for a CN-strong and a Na-poor RGB star, with the mean atmospheric parameters of their sample of NGC~6121 members, and they found a $\Delta(U-B)=0.04$~mag, that contributes to the observed dispersion in the cluster CMD.
In a later work, \citet{Milone2013} identified, through HST photometry, three distinct populations in NGC~6752 and they computed a set of synthetic spectra for two typical MS and RGB objects, combining the chemical mixtures of the three populations and different values of $Y$. They then computed the synthetic HST photometry to compare with the observed data and they conclude that both He and light-elements abundance variations should be taken into account to explain the observed colors.
In another study of NGC~6752, \citet{Dotter2015} produced a grid of several tens of model atmospheres and synthetic spectra, using two different codes: \textsc{ATLAS12/SYNTHE} \citep{Kurucz1979,Kurucz2005} and \textsc{PHOENIX} \citep{Hauschildt1999}. They explored a range of effective temperature ($3100 \lesssim T_{\rm eff} \lesssim 8000$~K) and surface gravity ($2 \leq \log{g} \leq 5.5$), suitable for the analysis of the MS and the base of the RGB, and two different mixtures for the abundances of light elements plus variations of C, N and He. They calculated the synthetic HST photometry, for all entries in their spectral library, to quantify the differences in colors and isochrones, both in the ultraviolet (UV) and near-infrared.
Additionally, as part of the {\em Hubble Space Telescope UV Legacy Survey of Galactic
GCs} \citep{Piotto2015}, \citet{Milone2015} present a multiwavelength photometric analysis that allowed to identify five distinct stellar populations in NGC~2808. They computed theoretical model atmospheres and synthetic spectra for a typical MS star for each population and found that   the HST colors in the UV and visible wavelength interval are very sensitive to changes in C, N, O, Fe, and He abundances.
\citet{Nardiello2015} also used HST photometry from a set of synthetic spectra to determine the He difference in the two populations of NGC~6352.
More recently, \citet{Milone2018} assessed the effects of He enhancement on the HST photometry by computing a large set of theoretical model atmospheres and spectra for stars in 57 GCs, also considering varying abundances of C, N, O, Mg, and Fe.
Finally, \citet{Milone2019} nicely reproduced the near infrared CMD of NGC~6752, obtained from HST observations, with synthetic photometry based on a set of theoretical spectra for representative stars of the three identified stellar populations of this cluster, placed along the MS over a large brightness range.

Aimed at complementing the chemically consistent isochrone computations \citep[e.g.][]{Salaris2006, Pietrinferni2009} and the need for better understanding the effects of abundance variations on photometric colors, \citet{Sbordone2011} conducted a theoretical stellar atmospheres analysis based on synthetic spectra computed for a set of four chemical element mixtures considered to be typical of subpopulations in galactic GCs. The goal of this paper is to extend the analysis of \citet{Sbordone2011}, under a similar framework, and explore the effects of the chemical variations on a set of spectroscopic indices that can provide the diagnostics of multiple stellar populations in GCs. 
We explore here which indices are especially appropriate for MS stars, as they have not yet suffered from convective chemical mixing that takes place in the RGB, so that the stellar surface still preserves the initial abundances; these objects are therefore suitable laboratories for studying the origin of the multiple population phenomenon in GCs. Nevertheless, we also provide indices for the analysis of the brighter RGB and HB stars.

In Sect.~\ref{sec:mix}, we describe the four chemical compositions from \citet{Sbordone2011}; then, in Sect.~\ref{sec:models}, we present the theoretical grids of model atmospheres and spectra, that we use to define and compute a set of spectroscopic indices (Sect.~\ref{sec:indices}), whose behaviour with atmospheric parameters is presented. The capacity of the indices of disentangling the effects of different stellar populations is quantified and commented in Sect.~\ref{sec:fiducialindices}, before a final summary.

\section{The chemical mixtures in GC\lowercase{s}}
\label{sec:mix}

For the calculation of synthetic spectra, we adopted the four chemical compositions of \citet{Sbordone2011}. They considered a reference isochrone representative for the First Generation population in a typical Galactic GC from the BaSTI data base\footnote{http://basti.oa-teramo.inaf.it/index.html} \citep{Pietrinferni2006}, with a He mass fraction of $Y$=0.246, a mass fraction of metals of $Z$=0.001, corresponding to an iron abundance of [Fe/H]=$-1.62$, with $\alpha$-elements enhancement of [$\alpha$/Fe]=0.4~dex and an age of 12~Gyr. This mixture is hereafter labeled as ``Reference", as in \citet{Sbordone2011}.

Two more metal mixtures representative of Second Generation (2G) stars in GCs were built using coeval isochrones calculated for the same reference parameters ($Y$=0.246, [Fe/H]=$-1.62$ and an age of 12~Gyr) but with different mixtures of C, N, O and Na that are characteristic of extreme values of the anticorrelations observed in galactic GCs \citep{Carretta2010}. The first of these mixtures, hereafter labeled as ``CNONa1Y2", includes N and Na enhancements of 1.8~dex and 0.8~dex by mass, respectively, and depletions of C and O by 0.6~dex and 0.8~dex, also by mass, respectively, with respect to the  Reference mixture. These conditions result in a $Z \simeq 0.00183$. This modification to the Reference mixture had already been proposed by \citet{Salaris2006}. The second of theses mixtures, labeled as ``CNONa2Y2", is the same as the CNONa1Y2 except that the enhancement of N is 1.44~dex by mass with respect to the Reference abundance (i.e. CNONa2Y2 is N-depleted with respect to CNONa1Y2). In this case, the metal distribution was constructed to keep $Z$=0.001.

Similar to CNONa1Y2, the fourth metal mixture for 2G stars, labeled as ``CNONa1Y4", considers an enhanced He mass fraction of $Y$=0.400. The Fe abundance and the age remain invariant in this chemical composition, while the global metallicity is $Z \simeq 0.00146$. The full table of elemental abundance of the four chemical compositions can be found in \citet{Sbordone2011} and, for the sake of easy reference, we show in Table~\ref{Chem} the abundances of the main elements considered in the four mixtures with respect to the Reference chemistry given in the first row of Table~\ref{Chem}, which provides the elemental abundances as $[\text{El}]_{\text{Ref}} = \log N(\text{El}) - log N(\text{H}) + 12$ for C, N, O and Na. The three modified mixtures are presented as the logarithmic difference of [El] with respect to $[\text{El}]_{\text{Ref}}$, 
while the He and the total metal content are expressed as the mass fraction $Y$ and $Z$.
It is important to stress that the differences in model atmospheres and synthetic spectra (see below), computed with the four chemical compositions, are caused not only by the variation of the light element abundances, but also by the change in the overall metallicity. This is particularly true when comparing the results obtained from CNONa1Y2 and CNONa1Y4, where the latter mixture has an enhanced He abundance but $Z$ is about 20 per cent lower, since the He content has observable lines only for hotter stars than those we consider in this work and, therefore, the effect of changing the He mass fraction on the stellar spectra is caused by a modification of the atmospheric structure.

\begin{table}
\caption{Relevant element abundance in the chemical mixtures.}
\begin{small}
\begin{tabular}{l c c c c c l }
\hline
Name      &  C     &  N     &  O    &  Na   &   Y     &  Z \\ \hline
Reference &  6.93  &  6.35  & 7.75  & 4.71  &   0.246 & 0.001  \\
CNONa1Y2  & -0.6   &  1.8   & -0.8  & 0.8   &   0.246 & 0.00183  \\ 
CNONa2Y2  & -0.6   &  1.44  & -0.8  & 0.8   &   0.246 & 0.001   \\
CNONa1Y4  & -0.6   &  1.8   & -0.8  & 0.8   &   0.400 & 0.00146  \\    \hline
\end{tabular}
\label{Chem}
\end{small}
\end{table}

\section{Calculation of ODF\lowercase{s}, atmospheric models and synthetic spectra}
\label{sec:models}

The full set of calculations presented in this section is based on the series of codes developed by Robert L. Kurucz \citep{Kurucz1970,Kurucz1979,Kurucz1992,Kurucz2005}. The first step in the computation of synthetic spectra is to build the means to take into account the opacity of tens of millions of atomic and molecular lines, while keeping the computational times reasonable. This is achieved by the so called opacity distribution functions (ODFs). While ODFs have been constructed for a variety of chemical compositions and other physical stellar parameters \citep[e.g.][]{Castelli2003}, they, however, do not consider the elemental partitions typical for GCs with one or more subpopulations.

For each chemical composition reported by \citet{Sbordone2011}, we generated ODFs tables using \textsc{DFSYNTHE}, a Fortran procedure written and upgraded by Robert Kurucz and made available by Fiorella Castelli at the Trieste astronomical observatory website\footnote{http://wwwuser.oats.inaf.it/castelli/sources/dfsynthe.html}. The required ancillary codes and procedures are described in \citet{Castelli2005b} to which the reader is referred for more details.

Subsequently, we implemented the ODFs as input in the \textsc{ATLAS9} code \citep{Kurucz1979,Kurucz2005} to compute a total of 3472 plane-parallel, LTE model atmospheres (868 for each chemical mixture) for a range in stellar parameters, namely: 
$T_{\text{eff}}$ from 4300 to 7000~K, at a step of 100~K, $\log{g}$ from 2.0 to 5.0~dex, with a step of 0.1~dex,
and the four chemical compositions of \citet{Sbordone2011}, described in Sect.~\ref{sec:mix}. The microturbulent velocity for all models was set to 2~km~s$^{-1}$, the standard value for \textsc{ATLAS9}-based model atmospheres.

\begin{figure}
\centering  
\includegraphics[width=0.45\textwidth]{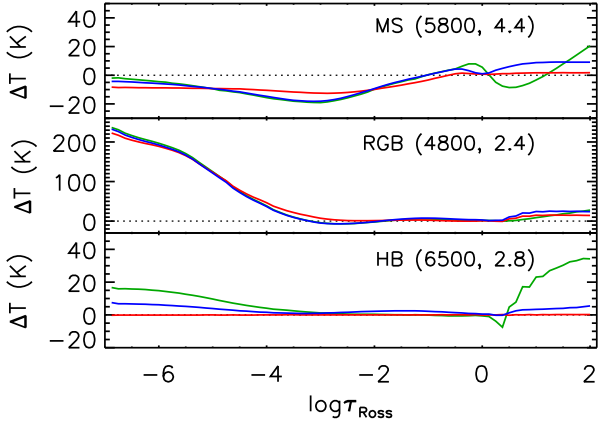}
\caption{Temperature profile difference of CNONa1Y2 (blue line), CNONa2Y2 (red line) and CNONa1Y4 (green line) models with respect to Reference mixture models as a function of $\tau_{\text{Ross}}$ for 3 representative models of different evolutionary stages, as indicated in the panel along with the $T_{\text{eff}}$ and $\log{g}$ of the model.}  
\label{models}
\end{figure}

In order to present how differing chemical partitions produce changes in the physical structure of atmospheric models, in particular in the temperature as a function of the Rosseland optical depth ($\tau_{\text{Ross}}$), we consider three representative models for MS, RGB and HB evolutionary stages, having ($T_{\text{eff}}$ (K), $\log{g}$ (dex)) = (5800, 4.4), (4800, 2.4) and (6500, 2.8), respectively. The results, shown in Fig.~\ref{models}, indicate that the biggest differences arise in the coolest models (up to $\sim 240$~K) in the outermost layers, while for warmer models the  discrepancy is far weaker and mostly located in the optically thick zone. These results are in accordance with those reported by \citet{Sbordone2011}.

Finally, for each model, a theoretical spectra at high-resolution ($R = \lambda/\Delta \lambda$=500,000) was computed using the \textsc{SYNTHE} series of codes, considering the wavelength interval 3300--6140~\AA, a spectral region that includes a large number of absorption lines and that is not significantly affected by chromospheric activity in the $T_{\text{eff}}$ range that we considered. The high spectral resolution will allow the use of the grid of synthetic spectra for a variety of observational data sets\footnote{The material presented in this section is available through a dedicated web page: \url{https://www.inaoep.mx/~modelos/gcindices.html}}. The analysis presented in what follows is based on a degraded version of the synthetic spectra to the intermediate resolution $R$=2500. We chose this resolution because it is compatible, for instance, with that of the OSIRIS spectrograph of the Gran Telescopio Canarias (GTC) in multi-object mode.

\section{Spectroscopic indices}
\label{sec:indices}
With the goal of identifying and analyzing promising spectral features that allow the identification of different stellar populations in GCs, we opted to use spectroscopic indices, as they have been implemented in the study of physical and chemical properties of astronomical objects in a number of astrophysical scenarios and with a diversity of objectives. In the particular case of the multiple population phenomenon in GCs, some works \citep[e.g.][]{Harbeck2003,Kayser2008,Pancino2010} have implemented the use of absorption indices for studying the chemical patterns. 

In this work, we have calculated from each of the degraded synthetic spectra the five indices reported by \citet{Pancino2010}, using the definitions provided in Sect.~3 of their paper; the indices are called S3839(CN), S4142(CN), CH4300, HK, and Hbeta.
However, the main goal of this work is to define a set of new indices that maximize the possibility of identifying stars that are members of different populations; this aim translates into maximizing the index differences, opportunely weighted, for the four different chemical compositions.
We describe the detailed procedure that we used to define the new indices in Appendix~\ref{appendix}. In brief, after visually identifying a promising spectral feature, we made a first guess of the wavelength limits of the index bands and we proceeded to modify these limits, using the Asexual Genetic Algorithm by \citet{Canto2009}, in order to maximize the absolute value of a merit function, that we called Error Weighted Difference ($EWD$):
\begin{equation}
\label{eq:ewd}
EWD = \frac{I_{\text{2}} - I_{\text{1}}}{\sqrt{\sigma_{\text{1}}^2+\sigma_{\text{2}}^2}} \, ,
\end{equation}
where $I_{\text{1,2}}$ are the index values for two different chemical compositions and $\sigma_{\text{1,2}}$ are the corresponding errors, obtained using a Montecarlo method with the assumption of a signal-to-noise ratio SNR=30 (at $\lambda$=4660~\AA), which corresponds to that expected in a $\sim 4$ hours exposure for spectroscopic observations of MS stars of M3, using OSIRIS at GTC. 
In addition to exploring the difference between the Reference chemical composition and any of the 2G mixtures (CNONa1Y2, CNONa2Y2, CNONa1Y4), we also consider those between 2G mixtures themselves: CNONa2Y2 vs. CNONa1Y2, for identifying indices sensitive to N change (and to the concomitant variation of $Z$), and CNONa1Y4 vs. CNONa1Y2, in search for indices sensitive to the He-enhanced composition, which is also  characterized by an increase in $Z$.

All our new indices have been computed using the same definition as the Lick/IDS indices \citep[see][]{Trager1998} and are expressed as pseudo-equivalent widths (in \AA):
\begin{equation}
\label{eq:index}
I = \int_{\lambda_1}^{\lambda_2} \left( 1 - \frac {F_{I}(\lambda)} {F_{C}(\lambda)} \right) d\lambda \, ,
\end{equation}
where $F_{I}$ is the flux in the feature passband, whose wavelength limits are $\lambda_1$ and $\lambda_2$, and $F_{C}$ is the interpolated pseudo-continuum (a straight line connecting the midpoints of blue and red passbands).

\begin{figure}
\centering  
\includegraphics[width=\linewidth]{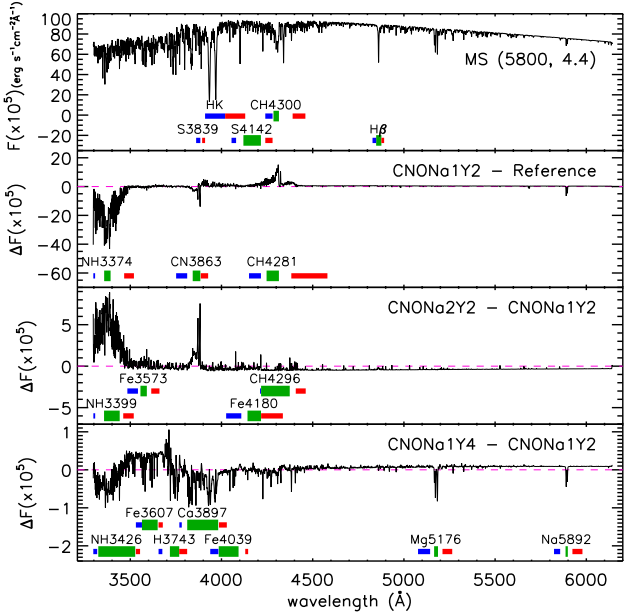}
\caption{Upper panel: the Reference spectra of the MS model (in parenthesis the $T_{\text{eff}}$ and $\log{g}$ values). The passbands of the five \citet{Pancino2010} indices are also plotted. Other panels: the flux difference between the MS model spectra with the chemical mixtures indicated in the panel. In these panels the passbands of the indices defined in this work are also displayed: in green the central passband, flanked by the blue and red sidebands.}  
\label{diffMS}
\end{figure}

\begin{figure}
\centering  
\includegraphics[width=\linewidth]{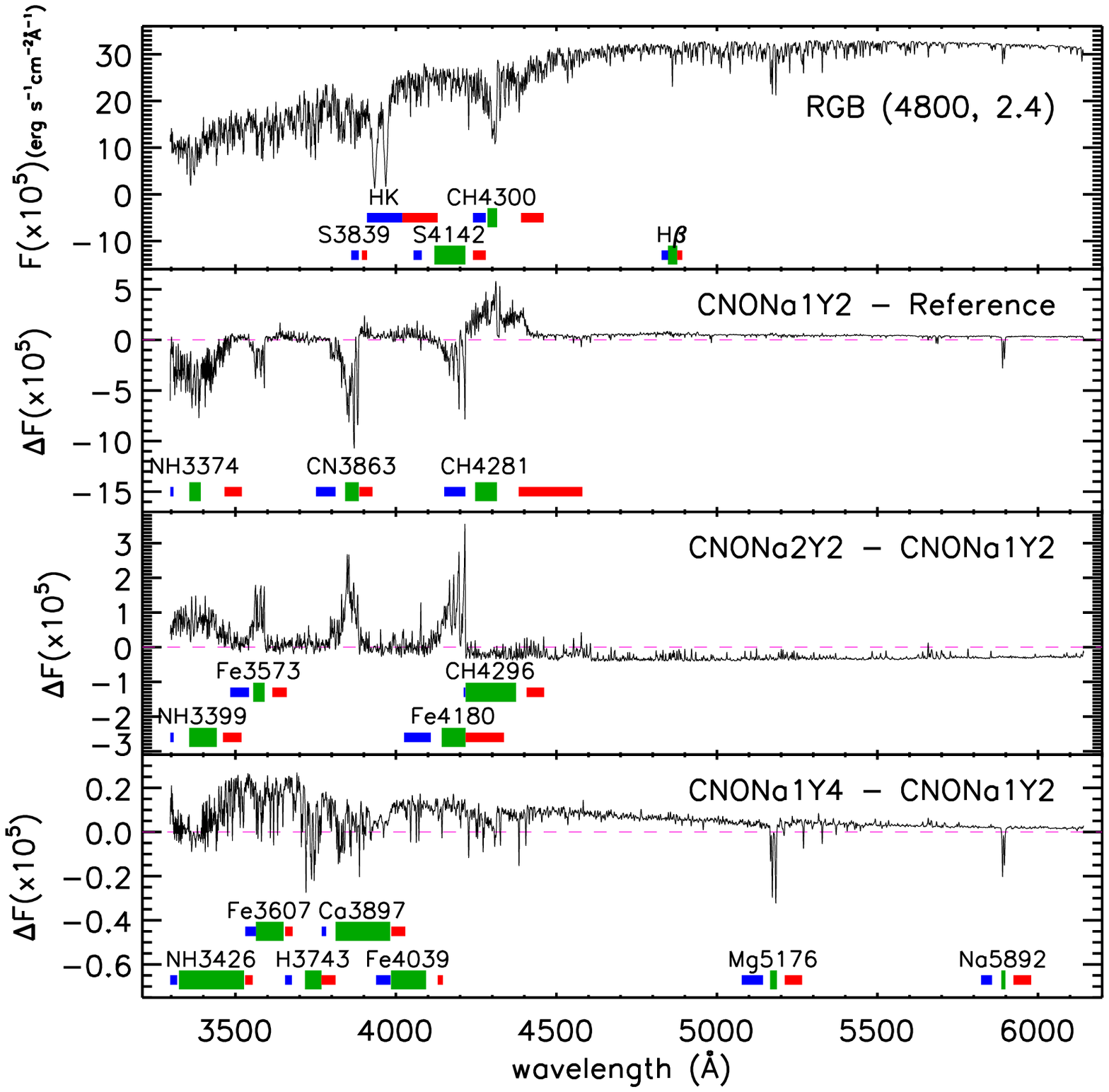}
\caption{Same as Fig.~\ref{diffMS}, but for the RGB model.}  
\label{diffRGB}
\end{figure}

\begin{figure}
\centering  
\includegraphics[width=\linewidth]{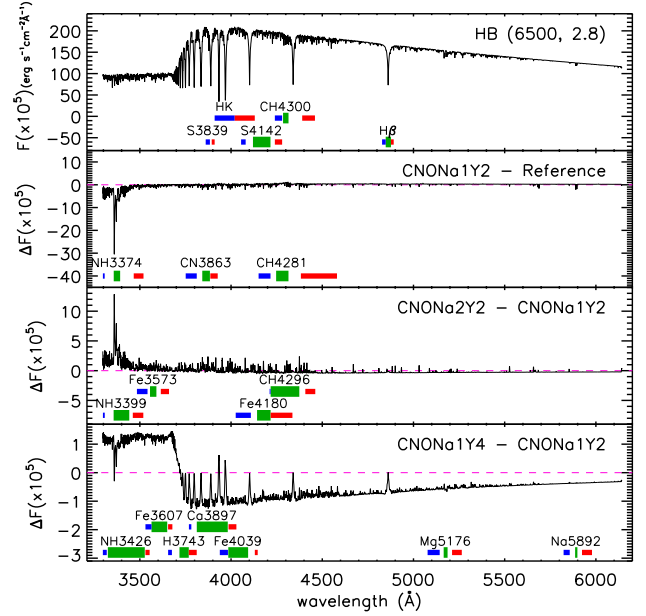}
\caption{Same as Fig.~\ref{diffMS}, but for the HB model.}  
\label{diffHB}
\end{figure}

The process culminated in a list of 14 indices, whose passband wavelength limits are reported in Table~\ref{Our_indices} (columns 3-5), along with the name (column 2; formed by the symbol of the dominant element and the midpoint of the central passband), a code (column 6) identifying the merit function used in the maximization process: ``2G" for the EWD between any of the 2G mixtures and the Reference one; ``N" for the EWD between CNONa1Y2 and CNONa2Y2; and ``He" for the EWD between CNONa1Y2 and CNONa1Y4. In column 7, we indicate the atomic or molecular species that mostly contribute to the absorption in the central passband, considering the MS representative model, and in parenthesis we add other relevant species that affect the side bands.
We present a graphical representation of the index passbands in Figs.~\ref{diffMS}, \ref{diffRGB}, and \ref{diffHB}.

\begin{table*}
\centering
\caption{Passband definition of the new spectroscopic indices defined in this work.}
\label{Our_indices}
\begin{tabular}{ r l c c c l l}
\hline
No. &  Name & Index passband [\AA] &   Blue passband [\AA] & Red passband [\AA] & Merit funct. &  Main species \\ \hline
 1 & NH3374 &   3356.7 --  3391.9 &    3297.4 --  3307.2  &   3466.4 --  3520.0 &  2G &  NH, Ti, Fe, Ni (Co) \\
 2 & NH3399 &   3356.1 --  3442.0 &    3297.5 --  3307.5  &   3461.3 --  3518.9 &  N  &  NH, Fe, Ti, Ni, Cr \\
 3 & NH3426 &   3324.6 --  3526.9 &    3297.1 --  3318.5  &   3530.5 --  3553.6 &  He &  NH, Fe Ti, Ni  \\
 4 & Fe3573 &   3555.8 --  3590.6 &    3484.0 --  3542.0  &   3615.0 --  3659.9 &  N  &  Fe, CN (Co, Ni)  \\
 5 & Fe3607 &   3563.9 --  3650.1 &    3530.7 --  3563.9  &   3654.7 --  3678.2 &  He &  Fe, Ni, Cr, Ti \\
 6 & H3743  &   3717.1 --  3768.2 &    3655.0 --  3675.7  &   3768.2 --  3811.7 &  He &  H, Fe, Ti (CN) \\
 7 & CN3863 &   3842.2 --  3884.1 &    3751.4 --  3812.0  &   3886.4 --  3926.8 &  2G &  Fe, CH, CN (H, Ca) \\
 8 & Ca3897 &   3812.2 --  3982.0 &    3768.9 --  3782.3  &   3986.1 --  4029.3 &  He &  Ca, CN, Fe, CH (H) \\
 9 & Fe4039 &   3984.8 --  4093.5 &    3938.4 --  3982.9  &   4130.3 --  4146.1 &  He &  Fe, Ti, H (Ca)  \\
10 & Fe4180 &   4142.5 --  4216.7 &    4025.6 --  4108.6  &   4217.3 --  4336.4 &  N  &  Fe, CH (Ti,H)  \\
11 & CH4281 &   4247.2 --  4314.7 &    4150.9 --  4216.3  &   4382.8 --  4580.7 &  2G &  CH, Fe, Ti, Ni (Co) \\
12 & CH4296 &   4217.1 --  4374.2 &    4211.4 --  4216.4  &   4407.1 --  4461.6 &  N  &  CH, H, Fe, Ti \\
13 & Mg5176 &   5165.7 --  5186.9 &    5077.4 --  5143.7  &   5211.6 --  5264.9 &  He &  Mg (Fe) \\
14 & Na5892 &   5886.1 --  5898.7 &    5823.1 --  5856.6  &   5923.9 --  5978.8 &  He &  Na (Fe) \\
\hline
\end{tabular}
\end{table*}

In the same figures, we also illustrate the spectral differences among the working chemical mixtures, for the three representative models for MS, RGB and HB stars, whose spectra are shown in the top panels. 
In the other panels, we show the flux residuals ($\Delta F$) between synthetic spectra of different chemical abundances, as labeled: in the second panel from top, the $\Delta F$ curve shows the effect of a 2G mixture, the third panel provides the effect of reducing the N abundance by 0.36 dex (and changing $Z$ from 0.00183 to 0.001), while the bottom panel presents the result of increasing He to $Y$=0.400 and decreasing $Z$ to 0.00146. 
In the top panel, we also show the passbands of the \citet{Pancino2010} indices, while in the other panels we depict the passbands of our new indices, grouped according to the merit function used to define them.
It is evident from the flux residuals that the main differences are present in the blue and near ultraviolet region of the spectral interval ($\lambda \lesssim 4500$~\AA): for MS objects, they are prominent around the CH feature at 4300~\AA\ (G band) and at $\lambda <$~3500~\AA, where the NH absorption produces the largest differences; it is also significant the combined effect of CN and CH on the blue side of the Ca~K line ($\sim$3840--3920~\AA). On the red side, Na lines originate the most significant flux differences, in particular the NaD doublet at 5890 and 5896~\AA, but also important is the region of the Mg triplet at about 5160~\AA\ for determining the effect of changing the atmospheric He abundance. In the case of RGB stars, the CN band at 3550--3600~\AA\ is also relevant for identifying multiple populations, while for HB members the spectral differences are more concentrated in the NH region, spiking at $\sim$3360~\AA. 

The spectral differences shown in Figs.~\ref{diffMS}, \ref{diffRGB}, and \ref{diffHB} are in overall agreement with those presented in previous works by \citet{Milone2013, Milone2015, Milone2018, Dotter2015}. Even though a quantitative comparison can not be carried out, as neither the chemical compositions nor the atmospheric parameters are the same, from a qualitative point of view all 
studies indicate that the major flux differences between spectra with chemical mixtures representative of different stellar populations arise at $\lambda < 4000$~\AA, where relevant absorption from molecules bearing light elements (mainly C, N, O) occurs.

In figure 11 of \citet{Milone2015} and in figure 4 of \citet{Milone2018}, the comparison of spectra of MS stars with different He abundance shows a significant flux difference, throughout the whole visible interval, that it is not present in our results (see the bottom panel of Fig.~\ref{diffMS}). In this case, one has to note that \citet{Milone2015} and \citet{Milone2018} compare stars with the same luminosity located on different isochrones and, therefore, with different $T_{\rm eff}$, that modifies the overall slope of the spectral energy distribution, while our match considers a constant $T_{\rm eff}$.

\subsection{Behaviour of the indices with respect to atmospheric parameters}

\begin{figure*}
\centering
\includegraphics[width=\textwidth]{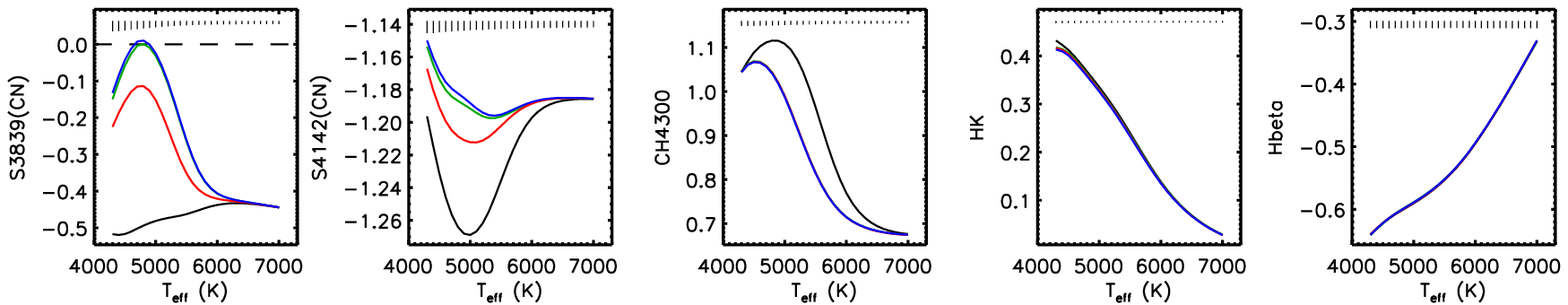}
\includegraphics[width=\textwidth]{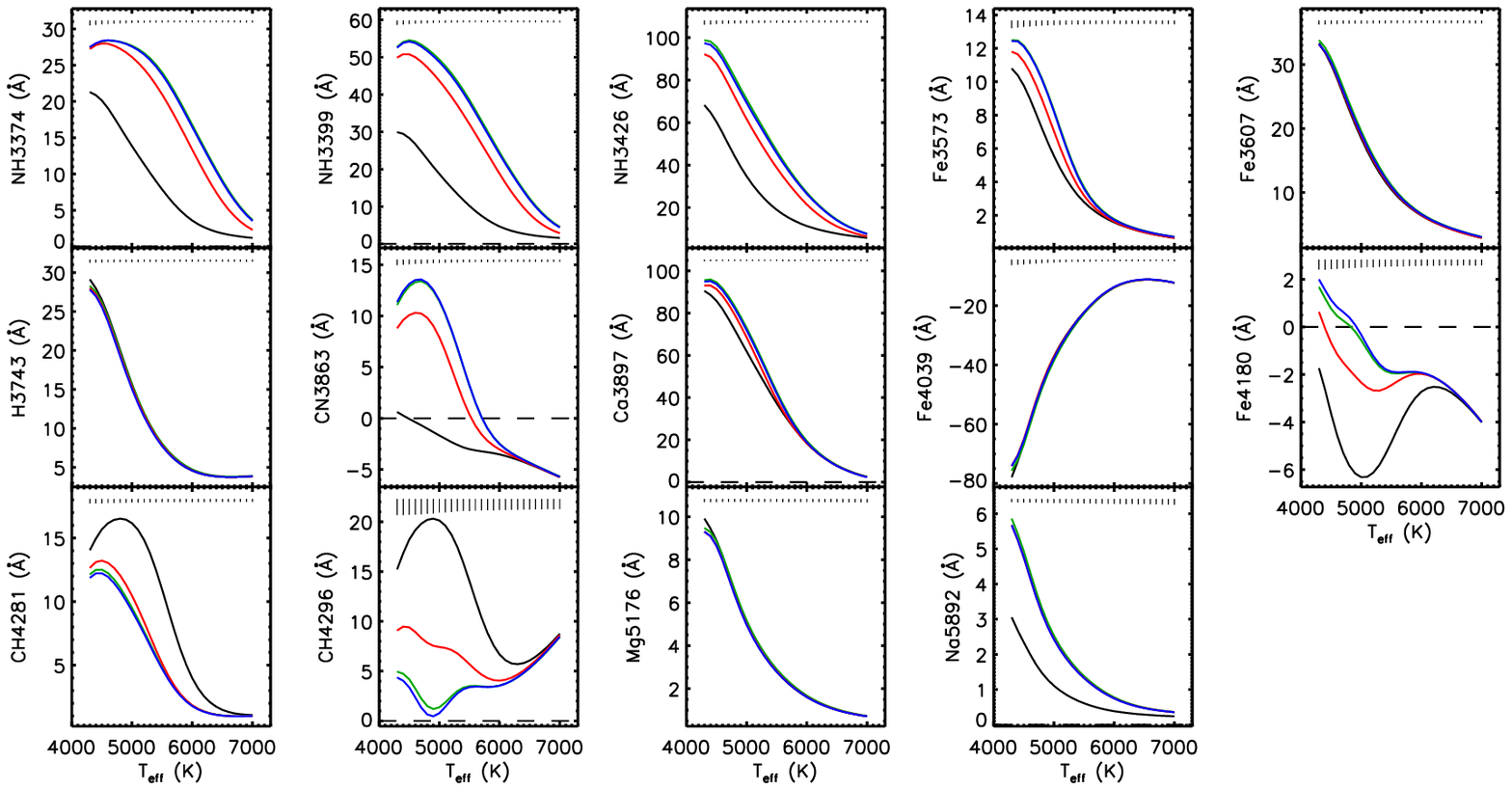}
\caption{Index values as a function of $T_{\text{eff}}$ for a fixed $\log{g}$=4.4~dex and for the Reference (black curve), CNONa1Y2 (blue curve), CNONa2Y2 (red curve), CNONa1Y4 (green curve) mixtures. The first row shows the \citet{Pancino2010} indices, while the following rows show the indices defined in this work. At each $T_{\text{eff}}$, the average of the index errors of the four mixtures, assuming a SNR=30, is indicated as a vertical bar near the top of each panel.}
\label{indicesteff}
\end{figure*}

Based on the calculations presented in the previous section, we present the results of the behaviour of the index values as a function of $T_{\text{eff}}$ for the different chemical mixtures. A visual inspection of the trends provides an indication of the parameter ranges where the indices might be useful for distinguishing subpopulations of GCs. 
Here we focus on the MS, so we present the results for models considering a fixed $\log{g}$=4.4, a typical value for cool stars on the H-burning evolutionary phase. Figure~\ref{indicesteff} shows the indices of \citet{Pancino2010} and those defined in this work. All indices were calculated for the four chemical mixtures which are portrayed with different line colors as indicated in the figure caption. The index variation with atmospheric parameters must be taken into account when building the index distribution of a sample of GC stars, that span a non-negligible range in those parameters, in search of a multimodal distribution that may point out the presence of different populations.

Regarding our indices, as we use a definition of the Lick/IDS indices, a positive value indicates that the flux in the central band is "in absorption" with respect to the pseudo-continuum, but, since the goal was to maximize the differences between chemical compositions, in some cases a low flux in the side passband(s) causes a negative index value.
In general, the indices absolute values show a decrement with increasing $T_{\text{eff}}$, because either these chemical species tend to be destroyed as the temperature increases or the lower level of the electronic transitions becomes less populated. A few indices do not follow this behaviour; this is due to the presence, inside or nearby the passbands, of one or more Balmer lines, whose absorption increases with increasing $T_{\text{eff}}$ in the range we consider here.

Indices where absorption of molecular species (CH, CN, NH) is relevant clearly differentiate the Reference chemical composition from the other mixtures, and they can also easily distinguish, amongst the 2G stars, the effect of changing N abundance. As the photospheric molecular content decreases at higher temperature, the indices values difference almost disappears at $T_{\text{eff}}$=7000~K, while it reaches a maximum for late-G or K-type stars.
Among 2G populations, the effect of the He-enhanced mixture is much smaller than that of the N-decreased one, but it is clearly noticeable in several indices, as in Fe4180, CH4296, and Pancino's S4142(CN); however, these three indices have a quite large error, suggesting that a deeper analysis is necessary to establish the significance of the differences caused by the CNONa1Y4 composition.

\begin{figure*}
\centering
\includegraphics[width=\textwidth]{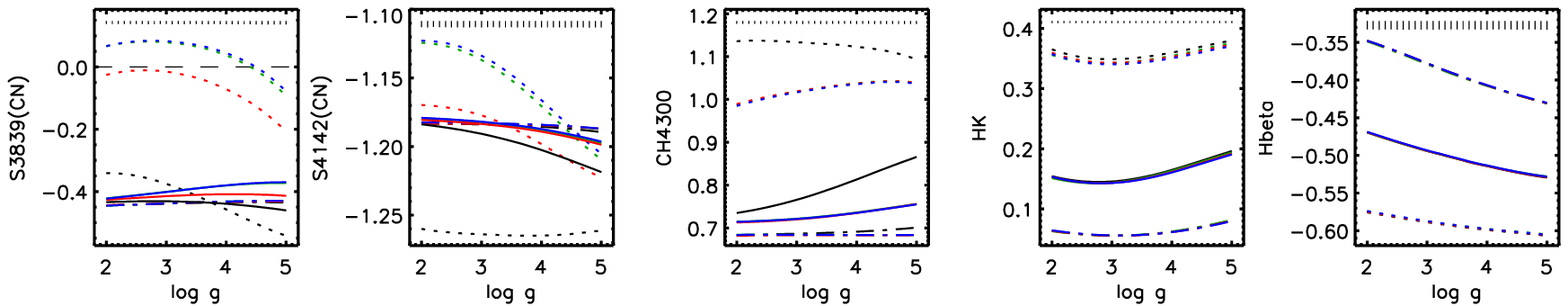}
\includegraphics[width=\textwidth]{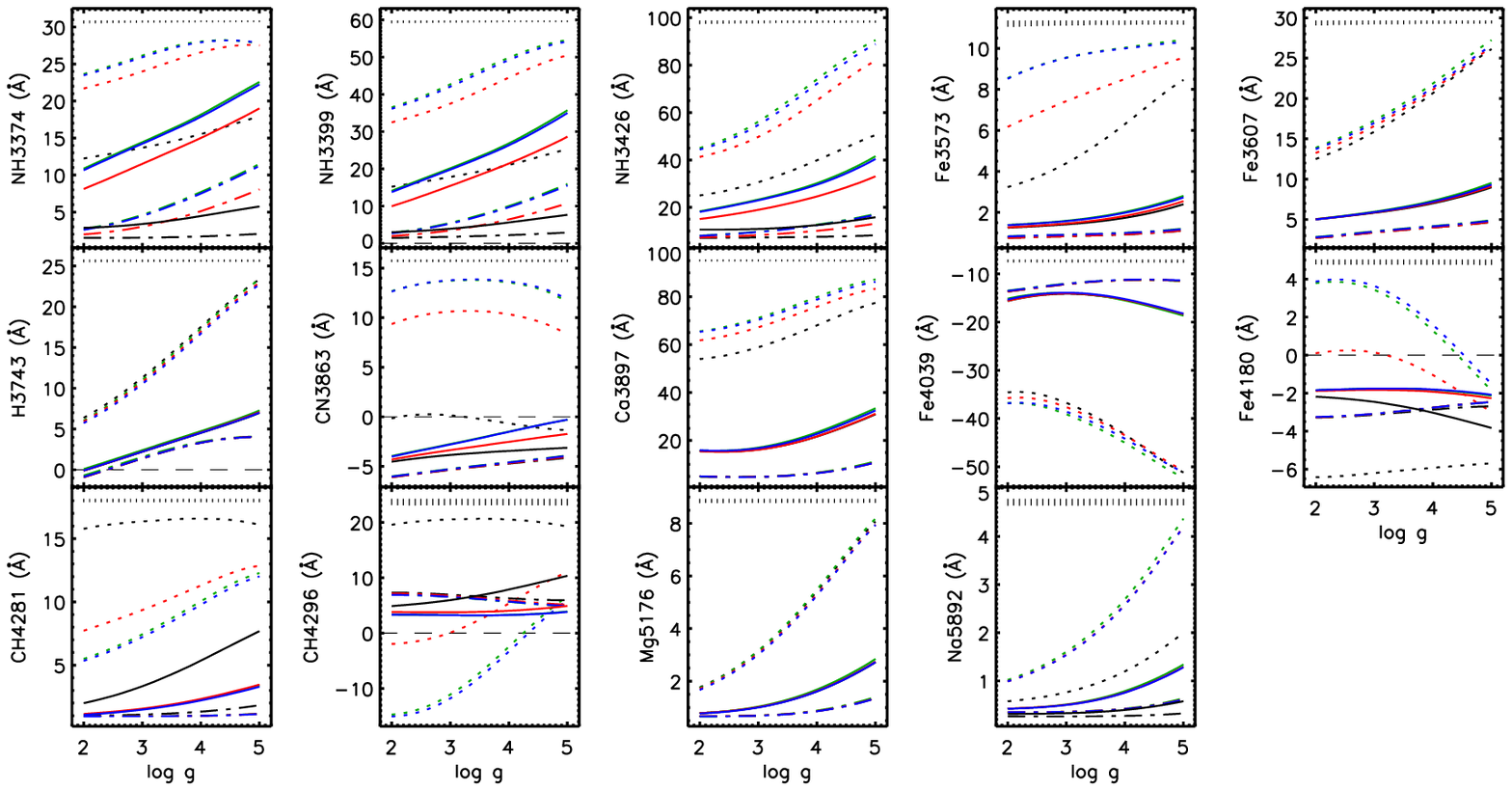}
\caption{Index values as a function of $\log{g}$ for 3 different $T_{\text{eff}}$ [4800~K (dotted curves), 5800~K (solid curves) and 6500~K (dash-dotted curves)] and for the Reference (black curves), CNONa1Y2 (blue curves), CNONa2Y2 (red curves), CNONa1Y4 (green curves) mixtures. The first row shows the \citet{Pancino2010} indices, while the following rows show the indices defined in this work. At each $\log{g}$, the average of the index errors of the four mixtures, assuming $T_{\text{eff}}$=5800~K and SNR=30, is indicated as a vertical bar near the top of each panel.}
\label{indiceslogg}
\end{figure*}

As shown in Fig.~\ref{indiceslogg}, a change of $\log{g}$, from 2.0 to 5.0~dex, also produces remarkable variations of the indices values. At a fixed $T_{\text{eff}}$=5800~K, as a general trend, the absolute value increases with increasing $\log{g}$, but also the difference between the Reference and the 2G indices increases, supporting the capacity of identifying 2G stars in the MS ($\log{g} \gtrsim 4.0$~dex). In the case of a fixed $T_{\text{eff}}$=6500~K, the change in $\log{g}$ does not produce large indices alterations, while towards the opposite end of the temperature interval (4800~K) the difference among different mixtures are significantly large and are, in most cases, larger at decreasing $\log{g}$.

\section{Fiducial indices for identifying multiple populations in GC\lowercase{s}}
\label{sec:fiducialindices}
The analysis of the indices that we illustrated in the previous section provides a qualitative indication on which features are the most promising for identifying and separating among different stellar populations. However, a quantitative study can be carried out by considering the \textit{EWD} and its dependence with $T_{\text{eff}}$. We recall that \textit{EWD} gives the difference between the values of two indices for different chemical composition in units of the typical errors ($\sigma$), therefore it strongly depends on the SNR and we remind that the results that are presented in this section are obtained by assuming SNR=30 at $\lambda$=4660~\AA.

\begin{figure*}
\centering  
\includegraphics[width=\textwidth]{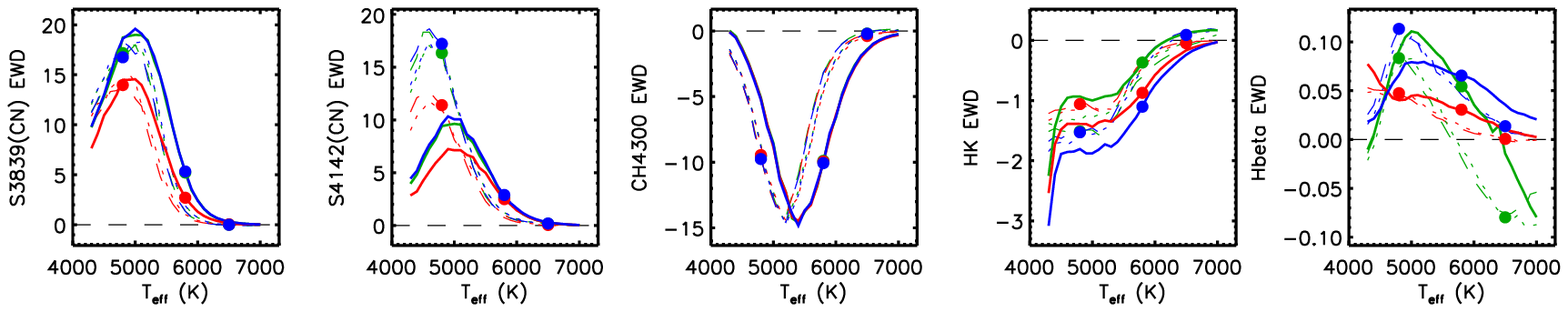}
\includegraphics[width=\textwidth]{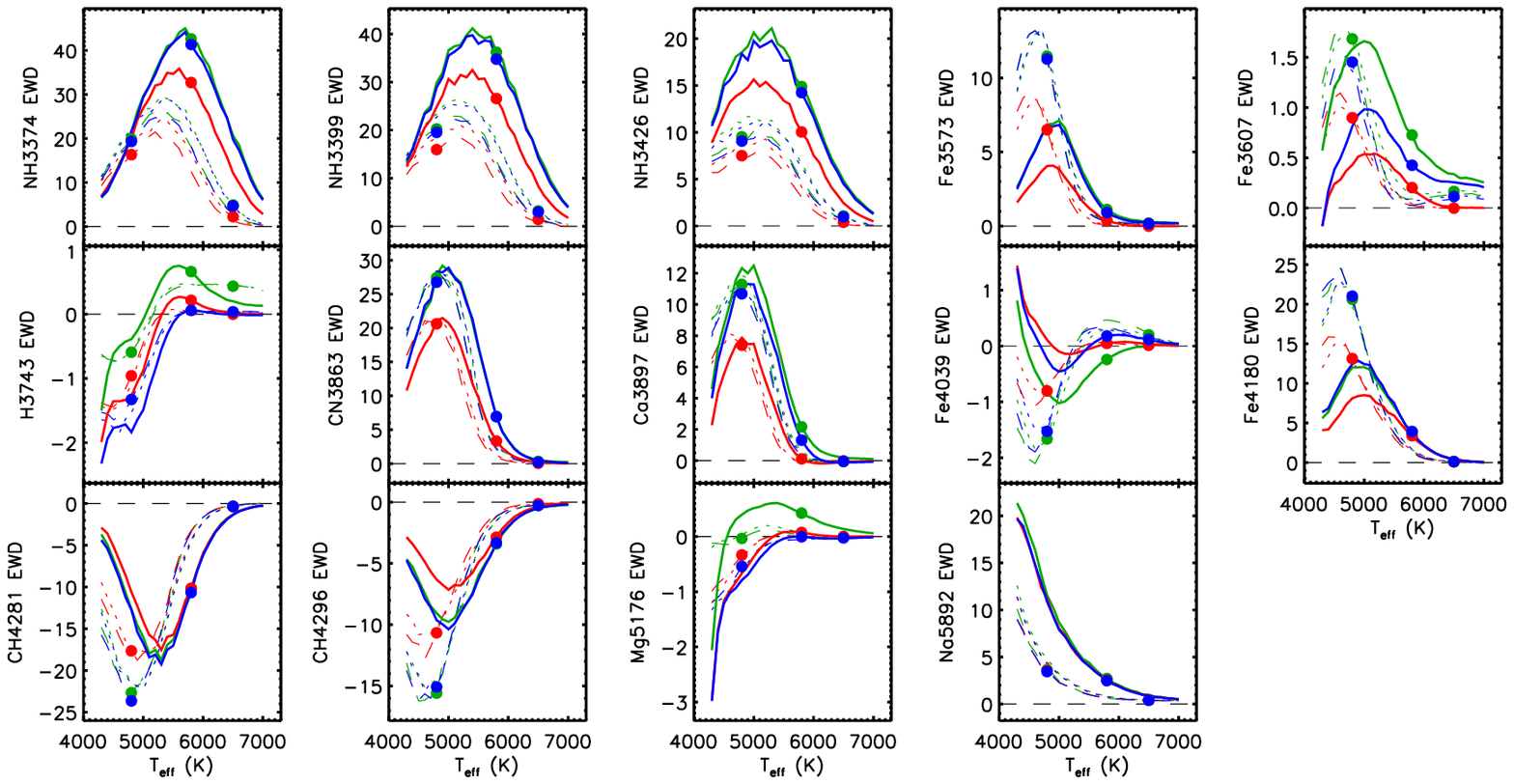}
\caption{EWD with respect to the Reference abundances as a function of $T_{\text{eff}}$ for the CNONa1Y2 (blue curves), CNONa2Y2 (red curves), and CNONa1Y4 (green curves) mixtures and for $\log{g}$=4.4~dex (thick solid curves), $\log{g}$=2.4~dex (dashed curves), and $\log{g}$=2.8~dex (dotted curves). SNR=30 is assumed. The first row shows the \citet{Pancino2010} indices, while the following rows show the indices defined in this work. The large dots mark the EWD of the representative models for MS, RGB and HB, with the same color code as the curves.}
\label{ewd}
\end{figure*}

\begin{figure*}
\centering  
\includegraphics[width=\textwidth]{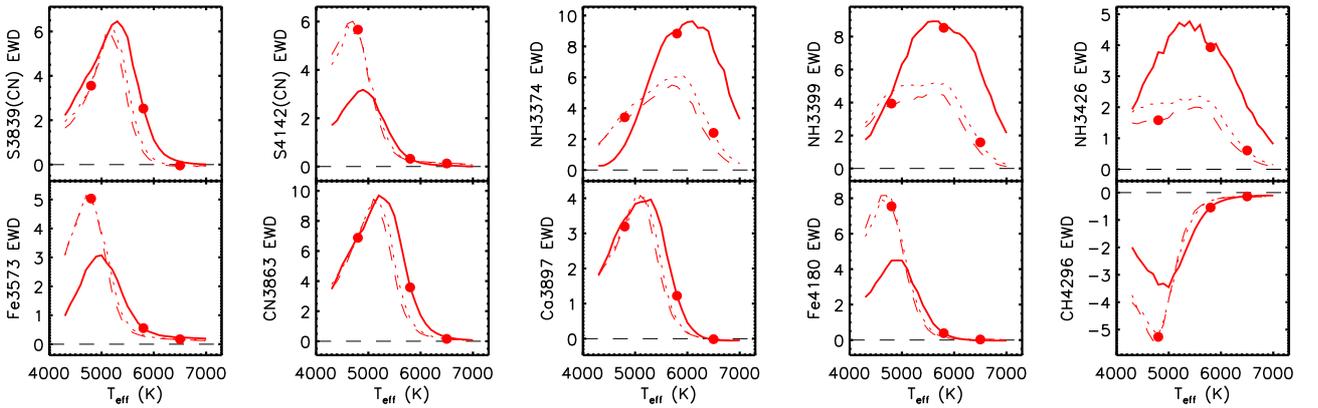}
\caption{EWD of CNONa2Y2 vs. CNONa1Y2 mixtures as a function of $T_{\text{eff}}$ for the same $\log{g}$ values as in Fig.~\ref{ewd} (SNR=30 is assumed). The first two upper left indices are from the \citet{Pancino2010} list. The filled circles mark the EWD of the models for MS, RGB and HB.}
\label{ewdnitrogen}
\end{figure*}

\begin{figure*}
\centering  
\includegraphics[width=\textwidth]{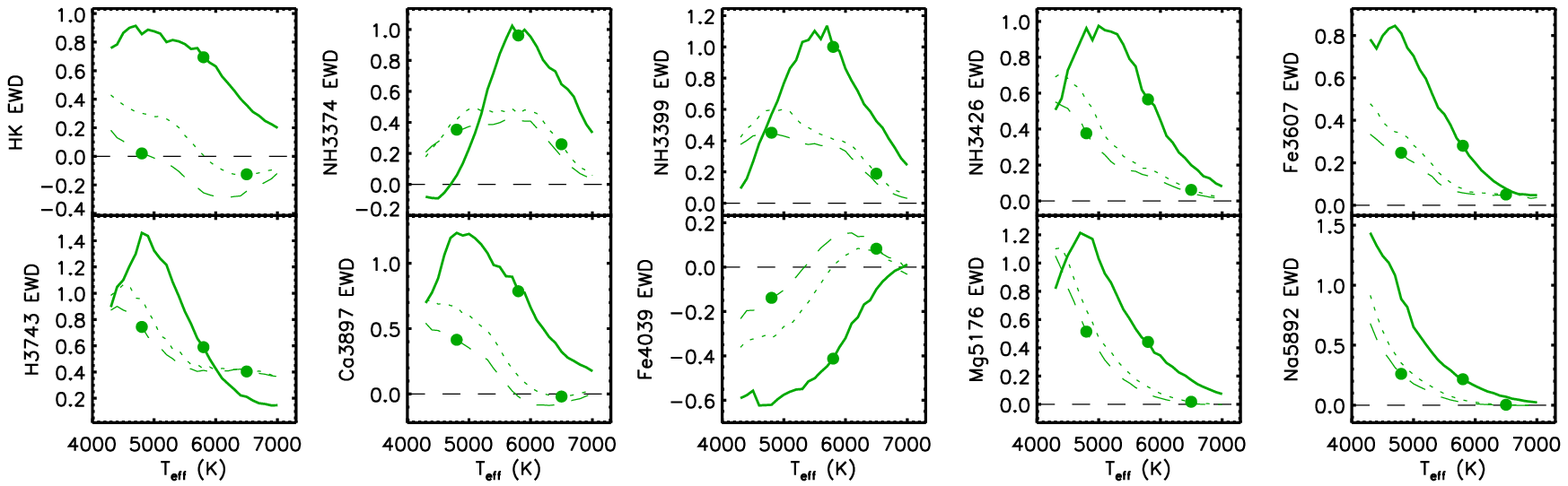}
\caption{EWD of CNONa1Y4 vs. CNONa1Y2 mixtures as a function of $T_{\text{eff}}$ for the same $\log{g}$ values as in Fig.~\ref{ewd} (SNR=30 is assumed). The first upper left index is from the \citet{Pancino2010} list. The filled circles mark the EWD of the models for MS, RGB and HB.}
\label{ewdhelium}
\end{figure*}

In Fig.~\ref{ewd}, we present, for all the indices, the \textit{EWD} between the three modified chemical compositions with respect to the Reference abundances as a function of $T_{\text{eff}}$. The \textit{EWD} can have both positive and negative values, but, as far as the capacity of differentiating between chemical mixtures is concerned, the absolute value is relevant and we adopt it for the following analyses. 
In order to better visualize the effect of the mixture with decreased N abundance,
in Fig.~\ref{ewdnitrogen} we plot the \textit{EWD} of CNONa2Y2 versus CNONa1Y2 compositions, but only for the subset of indices for which the effect is significant, while Fig.~\ref{ewdhelium} shows the \textit{EWD} caused by He-enhanced mixture (CNONa1Y4 versus CNONa1Y2) for the relevant indices.

The curves in the panels consider the $\log{g}$ values of 4.4, 2.4, and 2.8~dex, which are appropriate for MS, RGB, and HB  stars, respectively. We also add points (the filled circles) that mark the \textit{EWD} corresponding to the RGB ($T_{\text{eff}}$,~$\log{g}$)=(4800,~2.4) and HB (6500,~2.8) models, along with the MS (5800,~4.4) ones.

In general, all indices indicate that it is easier to identify stars with any of the 2G mixtures (Fig.~\ref{ewd}) than to distinguish among different 2G chemical compositions (Figs.~\ref{ewdnitrogen} and \ref{ewdhelium}). 
In particular, no index, at SNR=30, reaches a \textit{|EWD|}$>$3$\sigma$ when only the He-enhanced mixture is considered (Fig.~\ref{ewdhelium}); a quite larger SNR is therefore needed for disentangling stellar populations with this latter chemical composition at a statistical significant level.

A summary of the indices that are the most capable of distinguishing between objects with a Reference chemical composition and 2G mixtures and of discerning the effects of the mixtures with modified N or He abundance is presented in Table~\ref{bestindices}.
For these three cases, the table reports the \textit{|EWD|} values for the three models representative of the MS, RGB, and HB evolutionary phases. We considered a limit of 3$\sigma$  for the first two cases (labelled in the table as ``2G" and ``N"), but a much lower threshold of 0.40$\sigma$ for the sensitivity to He abundance change (labelled as ``He").
The first five rows present the \citet{Pancino2010} indices, followed by the new indices defined in this work.

\subsection{Pancino's indices}

Among the \citet{Pancino2010} indices, the most capable of identifying a 2G population in MS objects is CH4300, with a \textit{|EWD|}$\sim$10$\sigma$. However, the highest \textit{|EWD|} is reached by S3839(CN) and S4142(CN) at low $T_{\text{eff}}$, therefore this two indices are more suitable for detecting multiple populations in the RGB (both have \textit{|EWD|}=17.2$\sigma$). For this evolutionary phase, S3839(CN) and S4142(CN) also have a moderate capability of discerning N abundance variations.
The only \citet{Pancino2010} index with an appreciable sensitivity to the He-enhanced composition is HK, but only for MS stars.
The Hbeta index seems not at all useful for the chemical mixtures considered in this work.

\subsection{Our new indices}

\subsubsection{MS stars}
The most useful indices for detecting 2G objects in the MS are those that measure the strong NH absorption  at about $\sim$3350--3400~\AA: NH3374, NH3399, and NH3426. The first index reaches a record value of \textit{|EWD|}=42.6$\sigma$, while NH3399 is just a little less sensitive (36.2$\sigma$). These very large values indicate that these indices are still suitable when a quite lower SNR is achieved, considering, also, that the average flux value in the NH band is about a factor of 2--3 times lower than at $\lambda$=4660~\AA, where we defined the SNR for the error computation.
These two latter indices also are those with the highest sensitivity to the mixtures with modified N and He abundances: while they have \textit{|EWD|}$>$8$\sigma$ for the ``N" case, they reach just about $\sim$1$\sigma$ for the He-enhanced mixture, which means that a higher SNR should be attained to provide a statistically significant detection.
Examining the plots of NH3374, NH3399, and NH3426 in Figs.~\ref{indicesteff}--\ref{ewdhelium}, even tough the index values (i.e. the absorption) are larger at low $T_{\text{eff}}$ and $\log{g}$, the differences among the different chemical compositions are larger for MS G-type stars. This would make this set of indices ideal to study the stellar populations of the MS of GCs, considering that several spectrographs attached to large telescopes have the capability of observing the near-UV NH band\footnote{For instance, UVES and XShooter at the Very Large Telescope, HIRES at the Keck telescope, the Robert Stobie Spectrograph at the South African Large Telescope, the High Dispersion Spectrograph at Subaru, or MagE and MIKE at the Magellan telescopes.}.

Besides the NH case, the molecular bands of CN and CH (G band) are valuable for detecting multiple MS populations: the index CH4281 has a 10.7$\sigma$ sensitivity, but also CN3863, Fe4180, and CH4296 reach \textit{|EWD|}$>$3$\sigma$. However, if we exclude the mild sensitivity to CNONa2Y2 vs. CNONa1Y2 mixtures of CN3863 (3.6$\sigma$), these three indices are not useful for distinguishing populations with the chemical compositions with different N or He abundance. The index Ca3897, which includes in its central band both CN and CH lines and the
\ion{Ca}{ii}~HK 
lines, is quite sensitive to the He-enhanced (and larger $Z$) mixture. Since stars in Galactic GCs are old, this index should not be affected by chromospheric emission.

If we consider indices that measure spectral features with $\lambda \gtrsim 4000$~\AA, none of them show a significant sensitivity to the N abundance (i.e. CNONa2Y2 versus CNONa1Y2 mixtures) for for G2V or warmer stars and only Fe4180 and CH4296 could be of some use for objects of $T_{\text{eff}} \sim 5000$~K.
However, concerning He (i.e. CNONa1Y2 versus CNONa1Y4 mixtures), Fe4039 and Mg5176 have \textit{|EWD|}$>$0.4$\sigma$ for the MS model (5800,4.4), but many indices present an increasing sensitivity for cooler objects, reaching values larger than 1$\sigma$ for K dwarf stars. Again, these latter low \textit{|EWD|} values imply the necessity of having observations with SNR$>$30.

\subsubsection{RGB stars}

More indices appear appropriate for detecting different stellar populations using RGB objects than in the previous MS case, although no index reaches \textit{|EWD|}$>$30$\sigma$.

The three NH indices are, also in the RGB case, good choices for identifying 2G populations (\textit{|EWD|}$\gtrsim$10$\sigma$), but they are not the best ones. Furthermore, they have quite lower sensitivity to N-decreased and He-enhanced mixtures than for MS stars. 

The index Fe3573, whose central band is located in a region of densely packed metallic lines but that also includes strong CN absorption in cool objects, is useful for detecting 2G populations and it also has a good sensitivity to CNONa2Y2 versus CNONa1Y2 mixtures.

However, the best indices for RGB stars are those that measure stronger molecular bands of CN (at $\lambda$ $\sim$3810--3885 and $\sim$4140--4215~\AA) and CH (G band): CN3863 (that reaches \textit{|EWD|}=27.4$\sigma$), Ca3897, Fe4180, CH4281 and CH4296. 
Apart from Ca3897, they also show good sensitivity to the chemical composition with a modified abundance of N.

Near the red edge of the spectral interval, the index that measures the Na~D doublet (Na5892) is also marginally useful for detecting 2G stars.

The effects of the difference between CNONa1Y4 vs. CNONa1Y2 mixtures (the ``He" case) abundance are more difficult to detect in RGB than in MS objects: only four indices exceed the 0.4$\sigma$ threshold (NH3399, H3743, Ca3897, Mg5176). The most suitable one is H3743, especially defined for this case, that has \textit{|EWD|}=0.74$\sigma$.

\begin{table*}
\centering
\caption{\textit{|EWD|} values of the best indices for the representative models of MS, RGB, and HB stars.}
\label{bestindices}
\begin{tabular}{ r l c c c | c c c | c c c}
\hline
No. &  Name    & \multicolumn{3}{c}{2G} & \multicolumn{3}{c}{N} & \multicolumn{3}{c}{He} \\
    &          & MS   & RGB  & HB  &       MS & RGB & HB &          MS &  RGB & HB \\ \hline
 1 & S3839(CN) &  5.3 & 17.2 &     &          & 3.6 &    &             &      &      \\  
 2 & S4142(CN) &      & 17.2 &     &          & 5.7 &    &             &      &      \\
 3 & CH4300    & 10.1 &  9.8 &     &          &     &    &             &      &      \\
 4 & HK        &      &      &     &          &     &    &        0.69 &      &      \\
 5 & Hbeta     &      &      &     &          &     &    &             &      &      \\ \hline
     
 1 & NH3374    & 42.6 & 19.9 & 4.9 &      8.8 & 3.4 &    &        0.96 &      &    \\  
 2 & NH3399    & 36.2 & 20.2 & 3.2 &      8.5 & 3.9 &    &        1.00 & 0.45 &    \\
 3 & NH3426    & 14.9 &  9.5 &     &      3.9 &     &    &        0.57 &      &    \\
 4 & Fe3573    &      & 11.5 &     &          & 5.0 &    &             &      &    \\ 
 5 & Fe3607    &      &      &     &          &     &    &             &      &      \\ 
 6 & H3743     &      &      &     &          &     &    &        0.59 & 0.74 & 0.40 \\
 7 & CN3863    &  6.9 & 27.4 &     &      3.6 & 6.9 &    &             &      &      \\
 8 & Ca3897    &      & 11.3 &     &          &     &    &        0.79 & 0.41 &      \\
 9 & Fe4039    &      &      &     &          &     &    &        0.41 &      &      \\
10 & Fe4180    &  3.9 & 21.1 &     &          & 7.6 &    &             &      &      \\
11 & CH4281    & 10.7 & 23.6 &     &          & 5.1 &    &             &      &      \\
12 & CH4296    &  3.4 & 15.6 &     &          & 5.3 &    &             &      &      \\
13 & Mg5176    &      &      &     &          &     &    &        0.44 & 0.51 &      \\
14 & Na5892    &      &  3.6 &     &          &     &    &             &      &      \\
\hline
\end{tabular}
\end{table*}

\subsubsection{HB stars}
 
In the case of the HB model (6500,~2.8), the fact that most of the spectral features  tend to disappear at high $T_{\text{eff}}$ makes the indices of the different chemical compositions quite similar, so that they strongly diminish or completely loose the capability of discriminating among them. This behaviour also holds for those few indices where the value does not tend to zero at the warmer temperature.

In only three cases the \textit{|EWD|} is larger than the thresholds that we chose: NH3374 and NH3399 have \textit{|EWD|}$>$3$\sigma$ in the ``2G" case and H3743 just reaches \textit{|EWD|}=0.4$\sigma$, when the two mixtures that differ for the He content are considered.
The latter result suggest that, even though He-enhanced stars should populate the HB, a quite large SNR is required to recognize them from spectroscopic data.

\subsection{Indices dependence on SNR}

All the previous results have been obtained by assuming SNR=30 at $\lambda$=4660~\AA. We have conducted an exercise of computing \textit{EWD} for all indices and a number of SNR from a quite low value (SNR=10) to values that would require excessive integration times even for a 10-m class telescope (SNR=100). The result is that, for all indices and for all combinations of $T_{\text{eff}}$, $\log{g}$ and chemical composition, we found a clear linear relation between \textit{EWD} and the SNR, such that \textit{EWD}~$\propto \alpha$~SNR~+~$\beta$, where $\alpha$ is the slope and $\beta$ is the y-axis intercept. The errors on these parameters are very low and $\beta \approx 0$.

\begin{figure}
\centering  
\includegraphics[width=0.7\linewidth]{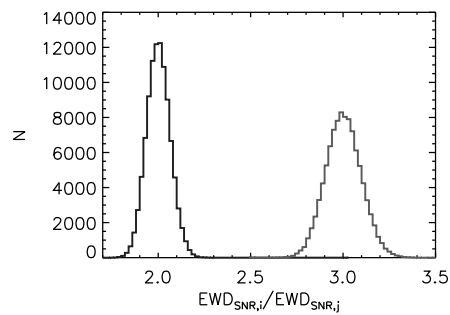}
\caption{Distributions of EWD computed at SNR=60 (left) and at SNR=90 (right) divided by the EWD computed at SNR=30, for all indices and for all combinations of $T_{\text{eff}}$, $\log{g}$ and chemical composition.}
\label{ewdsnr}
\end{figure}

A clear way of visualizing these linear relationships is presented in Fig.~\ref{ewdsnr}, where we plot the distribution of the ratio of the \textit{EWD} computed at SNR=60 divided by the \textit{EWD} computed at SNR=30, for all combinations of atmospheric parameters; if a linear relation passing through the origin exists, we should expect the ratio to be equal to 2. The distribution in Fig.~\ref{ewdsnr} has both mean and median values equal to  2.00, and the standard deviation (0.06) is caused by the random nature of the Montecarlo method used to compute the error. To further prove the point, we also display in Fig.~\ref{ewdsnr} the distribution of the ratio of the  \textit{EWD} computed at SNR=90 divided by the \textit{EWD} computed at SNR=30: also in this case the mean and median of the distribution have the expected value of 3.00.

\section{Summary}

We carried out a theoretical spectroscopic study in order to single out spectral features that allow the identification of multiple stellar populations in galactic GCs. We computed an extensive library of 3472 model atmospheres and high-resolution synthetic spectra using the collection of Fortran codes (\textsc{DFSYNTHE, ATLAS9} and \textsc{SYNTHE}) developed by R.~L.~Kurucz. The library includes four different chemical compositions, that are found in GCs with multiple stellar populations \citep{Sbordone2011}: one of them (labeled as Reference) is typical of stars of a First Generation, while the others, that consider abundance variations of several light elements such as He, C, N, O, and Na, are recognized as mixtures of 2G stars. 
The 2G chemical compositions vary among them for the N, He, and overall metal contents.
The grid of models and theoretical spectra covers a range of $T_{\rm eff}$  from 4300 to 7000~K and $\log{g}$ from 2.0 to 5.0~dex, suitable for the analysis of stars on the MS of GCs as well as other evolutionary stages, such as the RGB and HB. For each synthetic spectrum, that spans the wavelength interval 3300--6140~\AA, we calculated a total of 19 spectral indices: the five reported by \citet{Pancino2010} and 14 defined in this work that maximize the effect of the different chemical mixtures. The differences in chemical composition are quantified through the \textit{EWD}, which measures the difference between two indices in error units.

The spectral feature that provides the best capability (i.e. the highest \textit{EWD}) of identifying multiple population is the NH band at $\sim$3350--3400~\AA, which we probe with three different indices (NH3374, NH3399, and NH3426): it is most sensitive for objects in the MS, but is is also useful for RGB and HB stars. In fact, NH3374 and NH3399 are the only indices that reach a significant \textit{EWD} value for the hotter HB stars.
This NH band is also able to separate the two chemical compositions with changing N abundance (at least in the MS and RGB) and may also be sensitive to the He-enhanced mixture, if a high SNR is considered.

The G band at 4300~\AA\ is also useful for recognizing the presence of different stellar populations. Our three indices that incorporate the CH molecular band (Fe4180, CH4281, and CH 4296) are very sensitive to chemical differences between 2G stars and stars of the first generation, more so for objects in the RGB, for which they can also distinguish the N abundance. The intensity of the G band is also measured by one of the indices (CH4300) used by \citet{Pancino2010} and other authors, but it reaches lower \textit{EWD} than those provided by our indices.

The other profitable wavelength interval is 3800--4000~\AA, where, in addition to the \ion{Ca}{ii}~HK lines, molecular absorption of CN and CH is also relevant, at least for MS and cooler stars. We defined two indices that cover this wavelength range: CN3863 and Ca3897. The first one is useful for identifying 2G objects, both in the MS and in the RGB, while the second is only suitable for RGB objects. CN3863 also shows a mild sensitivity to N abundance, while Ca3897 may be useful for exploring the He atmospheric content.
The CN absorption is also measured by one of the \citet{Pancino2010} indices [S3839(CN)], but, again, it is less sensitive than those defined in this work.

Another of  our indices, Fe3573, is adequate to examine the chemical content of RGB stars, while the diversity between the two mixtures with different He content can be investigated by means of Fe4039, Mg5176, and H3743. The latter index shows a relative high \textit{EWD} not only for MS and RGB stars, but it is the only one that show some sensitivity in the HB. 

We expect that, through the appropriate combination of the spectral indices presented in this work, it would be possible to identify among stars belonging to different stellar generations in GCs.

\section*{Acknowledgements}

MC and EB would like to thank financial support from CONACyT grant CB-2015-256961. We also would like to thank the anonymous referee for her/his comments that helped to substantially improve this paper.

\bibliographystyle{mnras}
\bibliography{manuscript} 

\appendix

\section{Index definition}
\label{appendix}

We defined spectroscopic indices that are the most suitable for detecting differences in theoretical photospheric spectra of star with the same $T_{\rm eff}$ and $\log{g}$, but from model atmospheres of different chemical compositions, in the wavelength interval 3300--6140~\AA.
We adopted the index definition of the Lick/IDS indices \citep[see][]{Trager1998}, that make use of three passbands: a central one, where the index is measured, and two side bands (blue and red) that are used to determine a pseudo-continuum. The index value is computed by the Eq.~\ref{eq:index}. Therefore, the definition of an index implies the simultaneous optimization of six free parameters, that are the wavelength limits of the three passbands.

We tackle the problem using the  Asexual Genetic Algorithm  by \citet{Canto2009}. In brief, for each parameter, an initial random population of $N_0$ parameter values (i.e. individuals) is generated in a defined interval; a fitness value is associated to each individual by means of a merit function; a subset of $N_1$ individuals with the highest fitness is selected and passed to the next generation (for this reason the algorithm is called "asexual"); in a newly defined interval around each selected individual, a new generation of $N_2$ random individuals is created; the fitness of all individuals of the new generation is computed and the process is iterated until a stopping criteria is achieved. For more details on the algorithm, we refer to the paper by \citet{Canto2009}.

In this work, we proceeded as follows:

\begin{itemize}

\item We selected the $R$=2500 spectra of three combinations of $T_{\rm eff}$ and $\log{g}$, corresponding to the MS (5800, 4.4), RGB (4800, 2.4), and HB (6500, 2.8), and for the four chemical mixtures from \citet{Sbordone2011}, for a total of 12 spectra. We wanted to define a set of indices whose wavelength bands maximize the spectral differences between chosen chemical compositions.

\item By visual inspection of Figs.~\ref{diffMS}--\ref{diffHB}, we identified a promising spectral region and we made
a first guess of the wavelength limits of the three index passbands.

\item We generated a first generation of $N_0=186$ random wavelength values (individuals) in a $\Delta \lambda$=16~\AA\ interval centred at the first guess values. We imposed the constraints that the three passbands did not overlap and that each band were at least 5~\AA\ wide.

\item For each of the $N_0$ definitions, we computed the index value $I$ and the associated error $\sigma_{I}$ in all 12 stellar spectra. 
The error was obtained by using a Monte Carlo technique, similar to the one presented in \citet{Bertone2013}: at each $i$-th wavelength point in the passbands of the index, we added to the flux $F_i$ of the synthetic spectrum a flux error randomly extracted from a Gaussian distribution, centred in zero and with a standard deviation $F_i/(\text{SNR})$. We have assumed a  value of SNR=30 at $\lambda$=4660~\AA. We then computed the index value on this noisy spectrum and we repeated the procedure 500 times, obtaining a distribution of index values, whose standard deviation is considered as the typical error $\sigma_{I}$ of the spectroscopic index.

\item The index values $I$ and the associated errors $\sigma_{I}$ were used to compute the merit function, that we used for establishing the fitness of each individual, defined in Eq.~\ref{eq:ewd}, that we called \textit{EWD}.
The merit function was defined by adopting different combinations of chemical mixtures, when we wanted to define indices that were suitable for distinguishing: (i) 2G populations [\textit{EWD} from the difference of any of the 2G mixtures (CNONa1Y2, CNONa2Y2, CNONa1Y4) and the the Reference chemical composition], (ii) mixtures with varying N abundance (\textit{EWD} of CNONa2Y2 vs.~CNONa1Y2), or (iii) mixtures with different He content (\textit{EWD} of CNONa2Y2 vs.~CNONa1Y2).

\item For each parameter (each passband wavelength limit), we selected the $N_1=6$ fittest individuals, which produce the largest \textit{|EWD|} values. Around each of these individuals, we generated $N_2=30$ new random points in an interval  $\Delta \lambda_{\rm new} = \Delta \lambda_{\rm old}^{0.98}$, so that at each generation the wavelength interval became smaller (for instance, at iteration 28 $\Delta \lambda < 5$~\AA\ and at iteration 70 $\Delta \lambda < 2$~\AA). The values of $N_1$ and $N_2$ were defined so that the total size of each generation $N_0$ remained constant.

\item The process was iterated until the maximum wavelength difference among the $N_1$ fittest individuals was $\leq$0.5~\AA\, for each parameter, and the fittest individuals did not change for at least three iterations.
The six passband limits that provided the highest merit function were chosen for the index definition.

\end{itemize}

Since the merit quantity (\textit{EWD}) depends on the index error, which is obtained through a Montecarlo method, two runs of the genetic algorithm, with the same first guess inputs, provide slightly different results. We therefore carried out at least three runs for each index and chose the final passbands limits for the index definition among the best results of each run.


\bsp    
\label{lastpage}
\end{document}